\documentclass[journal]{IEEEtran}
\usepackage{graphicx,cite,amssymb,amsmath,multirow}
\usepackage{algorithm}
\usepackage{algorithmic}
\usepackage{multirow}
\usepackage{graphicx,subfigure}
\usepackage{epsfig}
\usepackage{bm}
\usepackage{subfigure}
\usepackage{array}

\begin{document}
%
\title{Optimal Time Scheduling Scheme for Wireless Powered Ambient Backscatter Communication in IoT Network}
\author{Xiaolan~Liu,
~Yue~Gao,  
~Fengye~Hu

\thanks{X. Liu and Y. Gao are with the Department of Electronic Engineering and Computer Science, Queen Mary University of London, London E1 4NS, UK (email: \{xiaolan.liu, yue.gao\}@qmul.ac.uk).

F. Hu is now with JiLin University, 130012, Changchun, China (email: hufy@jlu.edu.cn).

}

}

\maketitle
\begin{abstract}
In this paper, we investigate optimal scheme to manage time scheduling of different modules including spectrum sensing, radio frequency (RF) energy harvesting (RFH) and ambient backscatter communication (ABCom) by maximizing data transmission rate in the internet of things (IoT). We first consider using spectrum sensing with energy detection techniques to detect high power ambient RF signals, and then performing RFH and ABCom with them.
Specifically, to improve the spectrum sensing efficiency, compressive sensing is used to detect the wideband RF singals. We propose a joint optimization problem of optimizing time scheduling parameter and power allocation ratio, where power allocation ratio appears because REH and ABCom work at the same time. In addition, a method to find the threshold of spectrum sensing for backscatter communication by analyzing the outage probability of backscatter communication is proposed.
Numerical results demonstrate that the optimal schemes with spectrum sensing are achieved with larger transmission rates. Compressive sensing based method is confirmed to be more efficient, and that the superiorities become more obvious with the increasing of the network operation time. Moreover, the optimal scheduling  parameters and power allocation ratios are obtained. Also, simulations illustrate that the threshold of spectrum sensing for backscatter communication is obtained by analyzing the outage probability of backscatter communication.
\end{abstract}


\IEEEpeerreviewmaketitle


\section{Introduction}
\IEEEPARstart {T}{he} internet of things (IoT) is an intelligent network of different kinds of networks, which can connect various devices, smart sensors and actuators to the internet and enable information exchange and sharing among all IoT nodes\cite{A_vermesan2013internet}.
It connects a large number of nodes which are distributed in large areas within a complicated and heterogeneous environment to provide different applications such as, smart home, smart city and agriculture. Therefore, energy supply for such a large number of IoT nodes becomes a critical challenge, which indicating that these nodes have to achieve energy usage in a self-sustainable way. Energy harvesting is emerging as a promising way to provide nodes with continuous energy. Many research studies have confirmed that energy harvesting is a useful way to solve the problem of energy-limited batteries in wireless networks \cite{G_1kim2014ambient}, especially wireless energy harvesting\cite{H_kamalinejad2015wireless} which uses ambient radio frequency (RF) signals to achieve supplying power in IoT\cite{I_miorandi2012internet}. On the other hand, low power communication technologies have drawn some attention because they can extend the battery life of nodes by consuming less energy. For instance, low-power wide-area networks (LPWANs) which include LoRa, Sigfox and narrow band (NB)-IoT are regarded as promising communication technologies in IoT\cite{B_bor2016lora,C_lauridsen2017interference,D_beyene2017nb}. Specifically, ambient backscatter communication (ABCom), which is achieved by modulating ambient RF signals, is becoming popular due to its low power consumption characteristic.
ABCom doesn't contain energy-consumed circuits since it doesn't need the conventional transmitter circuit\cite{E_song2017internet}.

RF energy harvesting (RFH) and  ABCom have drawn much  attention because they consider readily available RF signals, such as signals from television (TV)/radio broadcasts, mobile base stations and handheld radios\cite{J_mishra2015smart} as available RF sources. These ambient RF sources can be classified into: static and dynamic ambient RF sources. TV signals are static RF sources since their power is relatively stable over time.
they are promising energy resources with their high transmission power (up to 10kW)\cite{K_harpawi2014design}. In\cite{L_keyrouz2012ambient}, the researchers designed a novel broadband Yagi-Uda antenna to harvest ambient RF power from DTV (Digital TV) broadcasting signals. A RFH wireless sensor network prototype was designed by harvesting signals from TV tower in \cite{M_nishimoto2010prototype}.
Although static RF sources can provide predictable RF energy, there could be long-term and short-term fluctuations due to service schedule\cite{ad1_lu2015wireless}.
Dynamic ambient RF sources (e.g., WiFi access point) are produced by RF transmitters that work periodically or change transmit power over time. Therefore, making use of ambient RF sources has to be adaptive and possibly intelligent to search for available opportunities in a certain frequency range.

Then, we first introduce the concept of spectrum sensing in this paper to detect the ambient RF sources, e.g., TV signals, by using energy detection techniques.
Spectrum sensing is first proposed to detect the signals sent by primary users in a licensed spectrum in cognitive networks\cite{S_zeng2010review}, which can provide more reliable and real-time results for spectrum occupancy. It is categorized into two types according to the bandwidth of the spectrum interest: narrowband and wideband spectrum sensing. The narrowband spectrum sensing algorithm includes energy detection, matched filtering and cyclostationary feature detection\cite{T_yucek2009survey}. Energy detection is most commonly researched due to its easy implementation and low computational complexity \cite{U_liang2008sensing}. But for exploiting the wideband spectrum, such as ultra high frequency (UHF) TV band, wideband spectrum sensing should be employed since narrowband sensing cannot be directly used due to its single binary decision making\cite{V_sun2013wideband}. Alternatively, compressive sensing (CS) theory, first proposed to apply to wideband spectrum sensing by Tian and Giannakis\cite{W_tian2007compressed}, could be used. Furthermore, a cognitive radio enabled TD-LET test-bed has been proposed in \cite{1_gao2016scalable}, authors use compressive sensing to detect TV white space to achieve dynamic spectrum management. CS uses a small number of measurements to reconstruct the received signals, and is popular due to low cost and high efficiency.

ABCom is emerging as a potential way to transmit information by reflecting an incident RF wave due to its characteristics of low power (tens of uW) and low complexity (without energy hungry circuits)\cite{N_kim2017hybrid}. Furthermore, ambient backscatter is used more to communicate with wireless devices nearby since it uses legacies of RF signals (TV, FM and WiFi signals)\cite{O_liu2017next}.
Leveraging the ambient RF signals contributes to a self-sustainable IoT network, that is, one which can use the RF signals to power the device and transmit information using the backscatter technique simultaneously\cite{P_han2017wirelessly}. Recently, ambient backscatter has been embedded into inexpensive objects to achieve the pervasive communication vision of IoT\cite{Q_shariatmadari2015machine}.
A communication system that enables two devices to communicate by leveraging existing TV signals was designed in reference\cite{R_liu2013ambient}. And reference\cite{add_1darsena2016achievable} has studied a general framework for evaluating the ultimate achievable rates of a point-to-point backscatter communication network.

In this paper, We first utilize the energy detection techniques in spectrum sensing to detect ambient RF sources with high power, and then use the detected RF signals to perform energy harvesting and backscatter communication. This makes the random energy arriving process become a deterministic process, and helps to find the best incident RF signal for ambient backscatter communication. We then propose optimal schemes to optimize time scheduling of different modules at IoT nodes by introducing spectrum sensing while maximizing the transmission rates of backscatter communication. Our results show that the optimal values of time scheduling parameters and power allocation ratios, and the maximum data transmission rates are obtained. Also, simulations demonstrate that the optimal scheme with a larger transmission rate is achieved with spectrum sensing, and the performance with compressive sensing is much better, and the superiorities become more obvious with the increasing of the network operation time. Then by analyzing the outage probability of backscatter communication with considering the channel suffering from path loss, shadowing and fading, we obtain the power threshold of spectrum sensing for backscatter communication.

This paper is organized as follows. The system model is built in section II. The optimal time scheduling schemes are proposed in section III. And analyzing the power threshold of spectrum sensing for backscatter communication in section IV. Numerical results corresponding to our analysis are shown in section V. Section VI gives the conclusions.

\section{System Model}
\begin{figure}[t]
\centering
\includegraphics[width=2.5in]{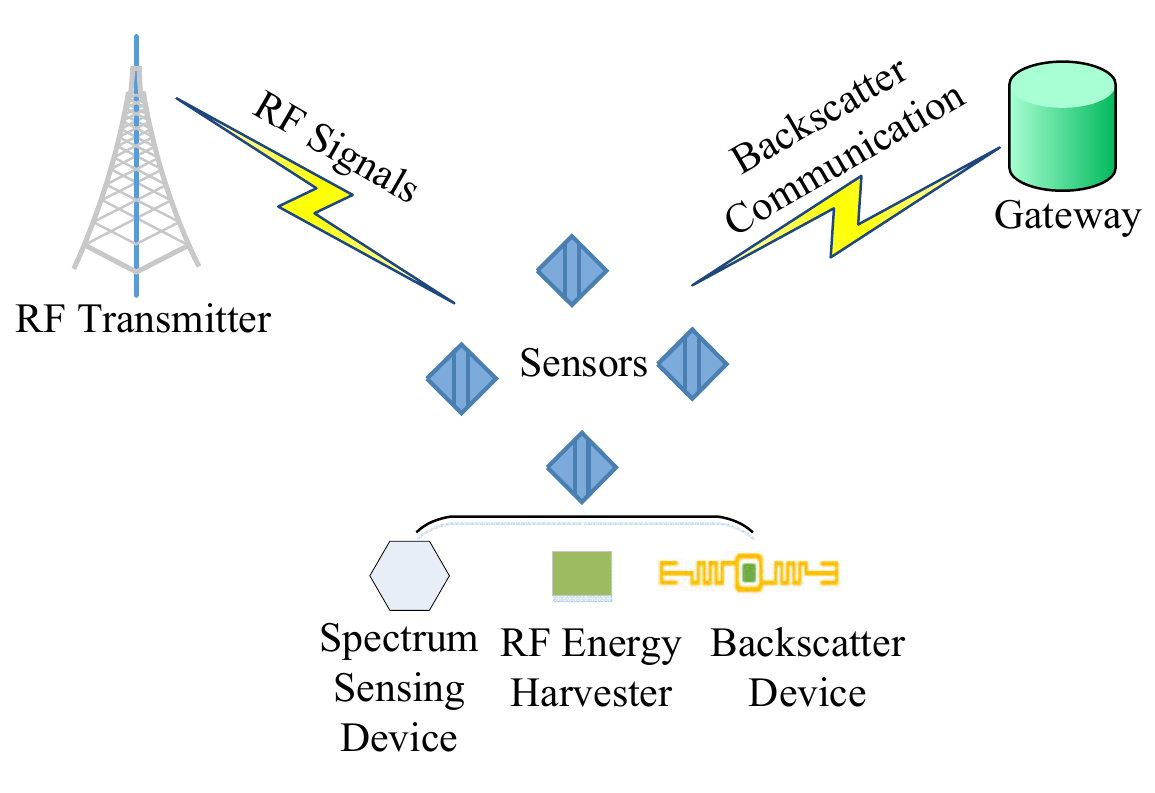}    
\caption{The diagram of network model} 
\label{fig1}                                 
\end{figure}

As shown in Fig.~\ref{fig1}, we consider a IoT network, it  contains a lot of IoT nodes which are powered by ambient RF sources and transmit data to the gateway by performing backscatter ambient communication using ambient RF sources.
Since ambient RF sources are existed almost everywhere around us, IoT nodes have abundant RF sources to perform RF energy harvesting and ambient backscatter communication. This general model can be used in many different fields, like smart city, smart home and smart agriculture. To make IoT nodes achieve energy self-sustainable and make the best of the ambient RF sources, we introduce RF energy harvester to harvest RF signals, spectrum sensing to detect ambient RF signals and ABCom to transmit data to the gateway. Certainly,  backscatter receivers are equipped at the end of the gateway to receive the data from the IoT nodes.

\begin{figure}[t]
\begin{center}
\includegraphics[width=3.3in]{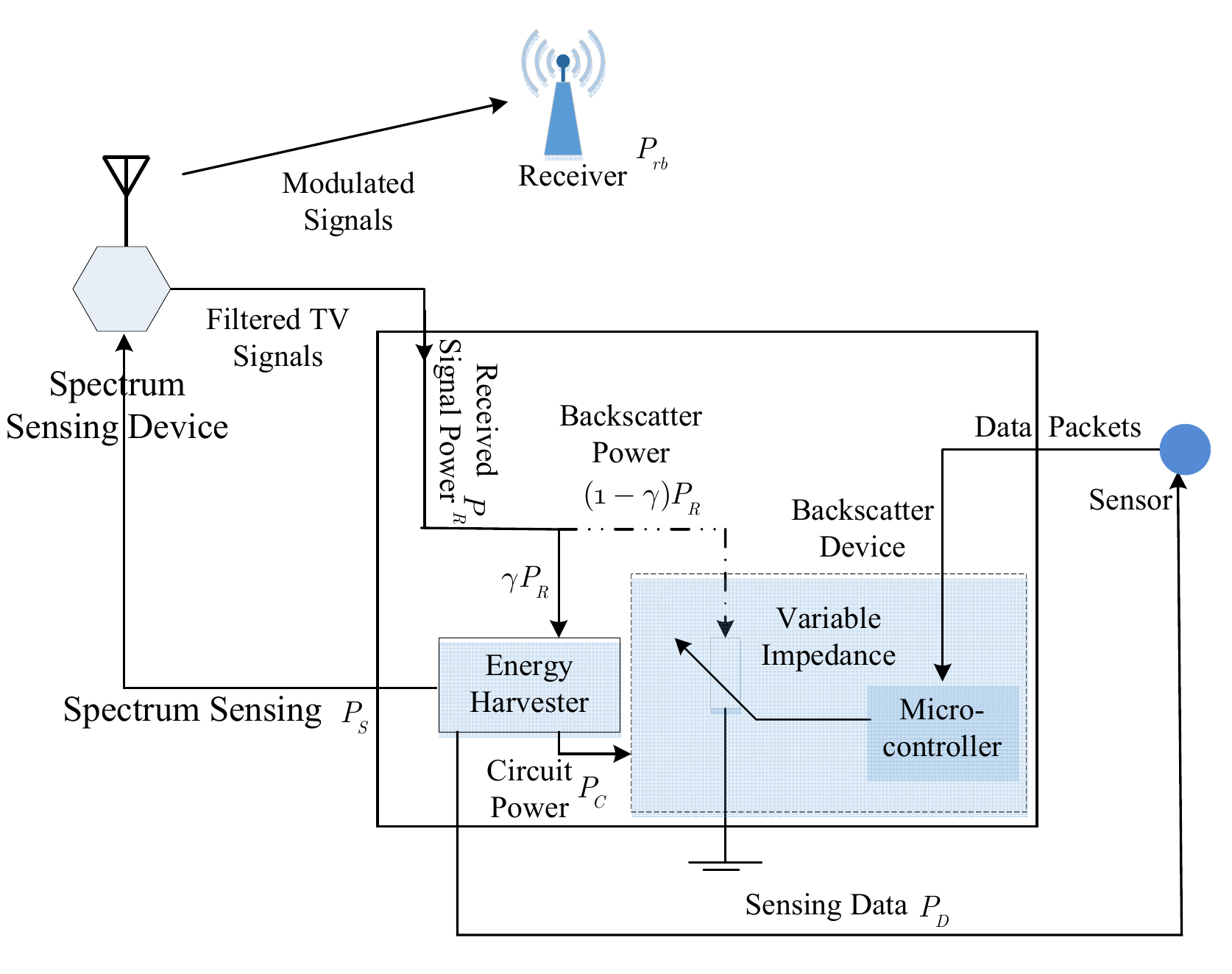}    
\caption{The process flowchart at the IoT node}  
\label{fig2}                                 
\end{center}                                 
\end{figure}
In Fig.~\ref{fig1},
to make the introduced three modules work properly at the IoT node, we have to design an optimal time scheduling scheme. The working process of all the modules at the IoT node is shown in Fig.~\ref{fig2}. The detailed process is formulated as: 1) The spectrum sensing module detects the ambient RF signals with high signal power. 2) The detected RF signals then be used to perform RF energy harvesting or ambient backscatter communication. 3) The energy harvester harvests the RF signals and then converts them into direct current (DC). 4) The backscatter device modulates data packets by using the detected high power RF signals, which is achieved by adjusting the variable impedance using the micro-controller.
In Fig.~\ref{fig2}, we can see that the harvested energy by the RF energy harvester is supplied to other modules, and we assume that the consumption energy by the RF energy harvester is negligible. So the energy causal condition should be satisfied, and it is expressed as:
\begin{equation}\label{eq-1}
{E_H} \ge {E_S} + {E_B} + {E_D},
\end{equation}
where ${E_H}$, ${E_S}$, ${E_B}$  and ${E_D}$  indicate the harvested energy, the energy consumed by the spectrum sensing module, the consumption energy of ABCom module and the energy consumed by the sensing module, respectively.
Assuming all the modules work in the same time block, then how to manage the time scheduling of different modules is a challenging probelm. So in this paper, the optimal time scheduling scheme is designed to solve this problem with IoT nodes achieving energy self-sustainable.

\section{Optimal time scheduling Scheme}

In this paper, the optimal scheme is proposed to manage the time scheduling of different modules at the IoT nodes based on spectrum sensing techniques. We adopt a saving-then-forward strategy, for a single frequency detecting, the detected RF signals ($100\%$) only can be either used for RFH by the RF energy harvester, or modulated ($100\%$) by the  ABCom  module. However, ambient RF sources are generally wideband signals, to detect all the ambient RF signals transmitted in a wideband spectrum range, compressive sensing technique is used. As a result, we can detect wideband frequency signals at the same time, and then the detected signals are used to perform energy harvesting (partial signals) and backscatter communication (the rest signals) simultaneously.

In \cite{M_nishimoto2010prototype}, the TV signals transmitted in the occupied primary channels can be harvested by the RF energy harvester, which is one of the important RF energy resources. In this paper, we consider TV signals as ambient RF sources, with frequency ranging from 470 to 790 MHz.
Even though TV signals are static RF sources, they could fluctuate over time due to service schedule.
Then the concept of spectrum sensing is introduced to detect the TV signals transmitted in the occupied channels. The energy detection technique, which is a classic method to detect the occupancy of primary channels by comparing the signal power of the received signals at the second user to a pre-defined power threshold. In this paper, we make some changes to the power threshold when perform energy detection for energy harvesting and backscatter communication. The threshold is set as the minimum energy value that can be harvested by the energy harvester successfully, while for backscatter communication, the threshold is set as a larger value to reduce the outage probability of data transmission.
And then we propose a method to determine the value range of the pre-defined threshold for backscatter communication by analyzing the outage probability of backscatter communication.

Assuming the TV signal received at the IoT node is $y\left( t \right)$, which is presented as
\begin{equation}\label{eq-2}
y\left( t \right) = x\left( t \right) + {n_o},
\end{equation}
where $x\left( t \right)$ is the received TV signal at the IoT node and $n_o$ is the noise.
Then the power of the received signal is calculated as
\begin{equation}\label{eq-3}
{P_R} = \frac{1}{N_s}\sum\limits_{n = 1}^{N_s} {{{\left| {y\left[ n \right]} \right|}^2}},
\end{equation}
where $N_s$ is the number of samples, $n$ is the index of discrete samples.
\subsection{optimal time scheduling scheme }
\begin{figure}[t]
\begin{center}
\includegraphics[width=2.5in]{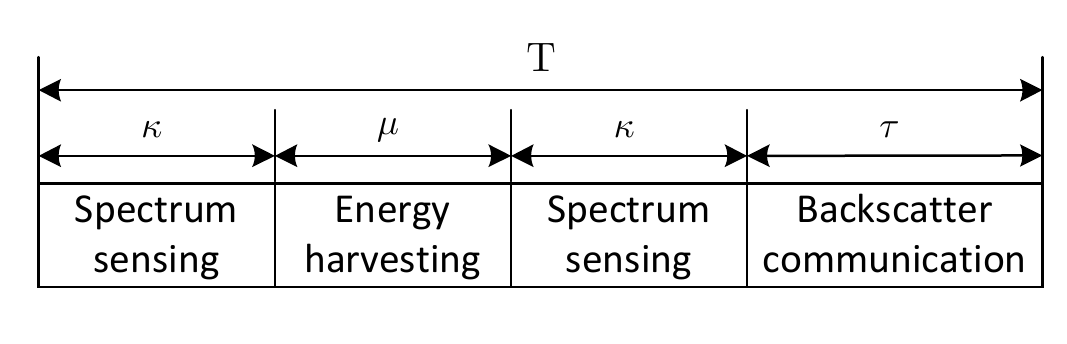}    
\caption{The time scheduling structure}  
\label{fig3}                                 
\end{center}                                 
\end{figure}

To design the optimal time scheduling scheme, we consider the energy causality at the IoT nodes, that is, the harvested energy must be higher than the consumption energy of other modules in a time block. As shown in Fig.~\ref{fig3}, the time scheduling structure presents the time block occupancy of each operation.
$T$ is the period of the time block, $\tau $ is the duration of backscatter communication, while $\mu $ is the duration of energy harvesting. The time duration of spectrum sensing is denoted by $\kappa$. In this case, spectrum sensing only detects a single frequency signal at a time, this means it has to detect TV signals every time before energy harvesting or backscatter communication.

Then each part of the energy causal condition in formula (\ref{eq-1}) is given respectively. The harvested energy by the RF energy harvester is given as
\begin{equation}\label{eq-4}
{E_{H}^n} = \mu T{\eta} {P_{R}^n},
\end{equation}
where $\eta$ is the energy harvesting efficiency and $P_{R}^n$ is the received signal power at the IoT node.

The consumption energy by the spectrum sensing module is presented as
\begin{equation}\label{eq-5}
{E_{\rm{S}}^n} = {e_s}{N_s}{M_n},
\end{equation}
where ${M_n}$ is the number of signals detected during spectrum sensing, and ${e_s}$ denotes the power consumed by each sample.

Assuming the pre-defined threshold for energy harvesting is ${\lambda _h}$, the RF energy harvester starts to harvest energy only when the detected signal power $P_{R}^n \ge {\lambda _h}$. And if the pre-defined threshold for ABCom module is assumed as ${\lambda _b}$, then it performs backscatter communication only when the detected signal power $P_{R}^n \ge {\lambda _b}$. Supposing the sampling rate of each signal is the same, denoted by ${f_s}$, so the total number of signals detected by the spectrum sensing module during $2\kappa T$ is ${M_n} = 2\kappa T{f_s}/{N_s}$. Thus, the energy consumed by spectrum sensing during one time block $T$ is changed as
\begin{equation}\label{eq-6}
{E_{\rm{S}}^n} = {e_s}2\kappa T {f_s}.
\end{equation}
And the consumption energy by the ABCom module is mainly used to power the ambient backscatter communication circuit, so it is expressed as
\begin{equation}\label{eq-7}
{E_{B}^n} = \tau T{P_C},
\end{equation}
where ${P_C}$  denotes the circuit power of the ABCom module. For simplicity, the consumption energy of sensing data ${E_{D}}$ is set as a constant.

To design the optimal time scheduling scheme, we consider the data transmission rate of backscatter communication as the objective function of the proposed optimization problem.
Then it is presented as
\begin{equation}\label{eq-8}
{R_n}(\tau ) = \tau TB_w{\log _2}(1 + \frac{P_U^n}{N_n}),
\end{equation}
where $P_U^n$ is the received signal power of the backscatter receiver at the data server. Since modulated signals by the backscatter module will experience path loss before arriving at the backscatter receiver, ${P_U^n}$ is written as ${P_U^n} = \beta {{P_{R}^n}}L(d_1)$.  ${P_{R}^n}$ is the received signal power after detecting by the spectrum sensing module, $\beta$ is the reflection coefficient of backscatter communication. In addition, it also suffers from path loss from the TV tower to the IoT node, which is denoted as ${P_{R}^n} = {P_o}L(d_2)$, $P_o$ is the power of TV signal at the transmitter. $L(d) = B{d^{ - \varsigma }}$ is the power-law path loss exponent. The path loss function depends on the distance $d$, a frequency dependent constant $B$ and an environment dependent path-loss exponent $\varsigma  \ge 2$.

Then the optimization problem is formulated as:
\begin{equation}\label{op1}
\begin{array}{l}
OP1\;\;\mathop {\max }\limits_\tau \;\; \;\;\;\;{{R_n}(\tau )}, \\
\;\;\;\;\;\;\;\;\;\;\;s.t.\;\;\;\;\;\;0 < \tau,\kappa,\mu < 1,\\
\;\;\;\;\;\;\;\;\;\;\;\;\;\;\;\;\;\;\;\;\;\;\tau+2 \kappa+\mu =1,\\
\;\;\;\;\;\;\;\;\;\;\;\;\;\;\;\;\;\;\;E_{H}^n \ge {E_{S}^n} + E_{B}^n + {E_{D}},
\end{array}
\end{equation}
where ${E_{D}} = {P_{D}}T$, $P_D$ is the power of the data sensing module. By substituting formulas (\ref{eq-4}), (\ref{eq-6}) and (\ref{eq-7}) into the optimization problem OP1, we can get the OP2 as follows:
\begin{equation}\label{op2}
\begin{array}{l}
OP2\;\;\;\;\mathop {\max }\limits_\tau  \;\;\;\;\;R_n(\tau ),\\
\;\;\;\;\;\;\;\;\;\;\;\;\;s.t.\;\;\;\;0 < \tau ,\kappa ,\mu  < 1,\\
\;\;\;\;\;\;\;\;\;\;\;\;\;\;\;\;\;\;\;\;\;\;\tau  = 1 - 2\kappa  - \mu ,\\
\;\;\;\;\;\;\;\;\;\;\;\;\;\;\;\;\;\;\tau  \le \frac{{\mu \eta {P_{R}^n} - {e_s}2\kappa {f_s} - {P_{D}}}}{{{P_C}}}.
\end{array}
\end{equation}
In the optimization problem OP2, we can see that ${R_n}(\tau )$ is a monotonically increasing function with $\tau $. In the ideal case, $\tau$ is as close as possible to 1, that is, $\kappa $ and $\mu$ are close to 0. However, RF energy harvesting process has to take up some time since the power harvested by it is small. Then we let $\kappa=0$, then $\mu  = 1- \tau$. Therefore, we can get the maximum transmission rate $R_n^*(\tau)$ when $\tau  = \frac{{\eta P_R^n - {P_D}}}{{\eta P_R^n + {P_C}}}$. In practical, spectrum sensing process cannot be negligible, that is, $\kappa\ne 0$. Then to solve this optimization problem with three variables, we use global search to obtain the optimal values.

\subsection{Timeslot Allocation Scheme with compressive sensing}
\begin{figure}[t]
\begin{center}
\includegraphics[width=1.8in]{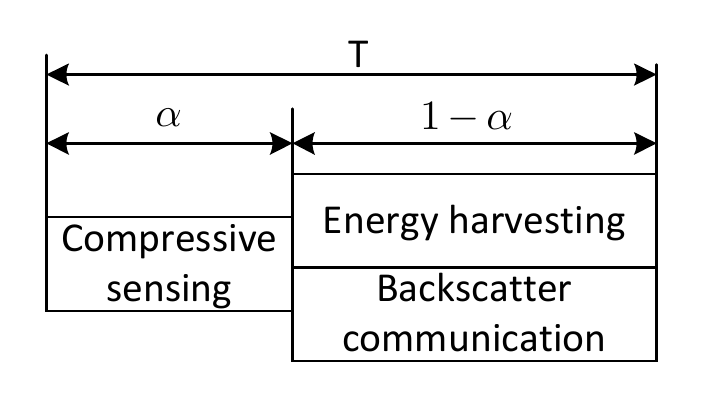}    
\caption{The time scheduling structure with compressive sensing}  
\label{fig4}                                 
\end{center}                                 
\end{figure}

In this section, compressive sensing is considered to detect wideband signals from all the TV channels ranging from 470MHz to 790MHz. Since compressive sensing has been proposed as a low-cost solution to reduce the processing time and accelerate the scanning process, we consider it as the main technique to detect the transmitted TV signals in primary TV channels. In this case, we can obtain a few TV signals from different channels at the same time.
The received power at the IoT node is divided into two parts, one is used to backscatter information, another part is harvested by the RF energy harvester, as shown in Fig.~\ref{fig2}. This means that energy harvesting and backscatter communication are performed simultaneously after compressive sensing. And the predefined threshold of energy detection is set as the bigger value of the two processes as defined in the previous time scheduling scheme.
In this case, the time scheduling structure is designed as shown in Fig.~\ref{fig4}. $T$ is the period of the time block. $\alpha$ denotes the time duration of compressive sensing, while $1 - \alpha $ indicates the time duration of energy harvesting and the backscatter communication.

Similarly, the harvested energy is used to supply power to the compressive sensing module, the ABCom module and the sensor module of the IoT node. Since only a part of the received power, which is indicated by power allocation ratio $\gamma$, is used for energy harvesting, the harvested energy is formulated as
\begin{equation}\label{eq-9}
{E_{H}^w} = (1 - \alpha )T\eta\gamma {P_{R}^w},
\end{equation}
where ${P_{R}^w}$ is the received signal power.

The consumption energy of the compressive sensing module is expressed as
\begin{equation}\label{eq-10}
{E_{S}^w} = {e_{s}}{f_s}{M_w}\alpha T,
\end{equation}
where ${e_{s}}$ is the consumption energy of each sample, ${f_{s}}$ is the sample rate of each signal. ${M_w}$ denotes the number of wideband signals, that is, the number of signals detected by compressive sensing.

The energy consumed by the ABCom module is presented as
\begin{equation}\label{eq-11}
{E_{B}^w} = (1 - \alpha )T{P_C}.
\end{equation}

In this case, the transmission rate of the backscatter communication is relevant to time scheduling of each module and the power allocation ratio between energy harvesting and  backscatter communication, then it is represented as
\begin{equation}\label{eq-12}
{R_w}(\alpha, \gamma ) = (1 - \alpha )TB{\log _2}(1 + \frac{{\beta(1-\gamma){P_{R}^w}}L(d_2)}{N_w}),
\end{equation}
where $(1-\gamma){P_{R}^w}$ indicates the part of received power used for backscatter communication.
$L(d_2)$ is the power-law path loss between the IoT node and the backscatter receiver.

Therefore, we formulate a joint optimization problem to optimize $\alpha$ and $\gamma$, which is given as
\begin{equation}\label{op3}
\begin{array}{l}
OP3\;\;\mathop {\max }\limits_{\alpha,\gamma} \;\;\;R_w(\alpha,\gamma ),\\
\;\;\;\;\;\;\;\;\;\;\;s.t.\;\;\;\;0\; < \alpha, \gamma  < 1,\\
\;\;\;\;\;\;\;\;\;\;\;\;\;{E_{H}^w} \ge {E_{S}^w} + E_{B}^w + {E_{D}^w}.
\end{array}
\end{equation}
Similarly, where ${E_D}$ is the energy consumed by the sensor sensing data. Substituting formulas (\ref{eq-9}), (\ref{eq-10}) and (\ref{eq-11}) into this optimization problem, it is reformulated as
\begin{equation}\label{op4}
\begin{array}{l}
OP4\;\;\;\;\mathop {\max }\limits_{\alpha ,\gamma } \;\;\;\;\;R_w(\alpha ,\gamma ),\\
\;\;\;\;\;\;\;\;\;\;\;\;\;s.t.\;\;\;0 < \alpha ,\gamma  < 1,\\
\;\;\;\;\;(1 - \alpha )\eta \gamma {P_{R}^w} \ge {e_s}{f_s}{M_w}\alpha  + (1 - \alpha ){P_C}+{P_{D}}.
\end{array}
\end{equation}
Likewise, the compressive sensing takes up as less time as possible in the ideal case, that is, $\alpha =0$. Then the joint optimization problem is simplified as the transmission rate only varies with the power allocation ratio $\gamma$. In addition, the transmission rate is monotonically decreasing with $\gamma$, so the optimal transmission rate $R_w^*(\gamma)$ is obtained with $\gamma  = \frac{{{P_C} + {P_D}}}{{\eta P_R^w}}$. However, in practical, time slots performing compressive sensing cannot be negligible, then we use global search to solve this optimization problem.

\section{Optimal ABCom Threshold}
In Section III, we have proposed the optimal scheme to manage time scheduling of different modules at IoT node. In the optimal scheme, spectrum sensing technique is first used to detect the transmitted TV signals instead of the vacant frequencies as conventional. The signals are detected by changing the power threshold of energy detection technique, which then are used to perform energy harvesting and backscatter communication. The pre-defined threshold for energy harvesting is set as the minimum value that can be harvested by the RF energy harvester. But for backscatter communication, the pre-defined threshold is decided by the interferences of other IoT nodes and the channel condition between backscatter device and the receiver. Then in this section, we will discuss the upper and lower bound of the pre-defined threshold for backscatter communication.

As shown in Fig.~\ref{fig1}, the IoT network contains a lot of IoT nodes distributed around the gateway (which is considered as the local data server). If these nodes transmit data with high transmission power, it will cause interferences to each other. It means the transmission power of each IoT node should be no more than a maximum power. So the upper bound of the pre-defined threshold is existed. It is obtained by maximizing the transmission rate of backscatter communication. For simplicity, we assume the transmission power of each IoT node is the same, then the transmission rate is presented as
\begin{equation}\label{eq-13}
\begin{array}{l}
{R_b} = B{\log _2}(1 + \frac{{{P_l}h}}{{{N} + \sum\limits_{k = 1,k \ne l}^K {{P_k}{g_k}} }})\\
\;\;\;\;\; = B{\log _2}(1 + \frac{{{P_l}h}}{{{N} + (K - 1){P_l}\sum\limits_{k = 1,k \ne l}^K {{g_k}} }}),
\end{array}
\end{equation}
where $N$ is the noise of the receiver, and $P_l=P_k$ is the transmission power of IoT node. $h$ is the channel gain from backscatter device to the receiver, $g_k$ is the channel gain from the rest IoT nodes to the IoT node $l$. And as shown in formula (\ref{eq-13}), the transmission rate $R_b$ is monotonically increasing with the transmission power $P_l$. Generally, the upper bound of the transmission power is limited by the capacity of the battery at the IoT node.
In this paper, since the backscatter device transmits data by reflecting the transmitted TV signals, the upper bound of transmission power is decided by the maximum power of the TV signals detected by the spectrum sensing module.

Then we discuss the lower bound of the pre-defined threshold which ensures successful data transmission from the backscatter device to the receiver. In reality, the channel between them suffers from large-scale fading (including path loss and shadowing) and small-scale fading (like Rayleigh and Nakagami-m fading).
In this paper, we consider this channel with all the fading characteristics to find the transmission power threshold of the backscatter communication, that is, the pre-defined threshold of spectrum sensing for backscatter communication can be obtained.

Firstly, we use the simplified path loss model to illustrate the influences of the communication distance on the transmission power, it is expressed as
shown in Section III.A. So we can get
\begin{equation}\label{eq-14}
{P_t} = \frac{{{P_{R}^b}}}{{B{d^{ - \varsigma }}}} \ge \frac{{{P_{th}}}}{{B{d^{ - \varsigma }}}},
\end{equation}
where $P_{th}$ is the threshold of the received power at the receiver. Then we can use formula (\ref{eq-14}) to model the path loss of the signals before they arrived the receiver. Then it is easy to obtain the minimum transmission power according to the received power that should exceed a minimum threshold to make sure the successful data transmission.

Then, we consider finding $P_{th}$ by analyzing the outage probability of the channel with composite Nakagami-m fading and log-normal shadowing. This is to analyze the effects of shadowing and fading in the channel on the received power. Firstly, the definition of the outage probability is presented as
\begin{equation}\label{eq-15}
{P_{out}} = {{\mathop{\rm P}\nolimits} _r}(\gamma  < {\gamma _{th}}) = \int_0^{{\gamma _{th}}} {{p_\gamma }(\gamma )} {d_\gamma },
\end{equation}
where $\gamma$ is the instantaneous signal-noise ratio (SNR) and $\gamma_{th}$ is the minimum SNR that must be satisfied at the receiver.
Then, the composite probability density function (PDF) with Nakagami-m fading and log-normal shadowing of SNR is presented as\cite{add_2simon2005digital}
\begin{equation}\label{eq-16}
\begin{array}{l}
{p_\gamma }(\gamma ) = \int_0^\infty  {\frac{{{m^m}{\gamma ^{m - 1}}}}{{{\Omega ^m}\Gamma (m)}}} \exp ( - \frac{{m\gamma }}{\Omega })\\
\;\;\;\;\;\;\; \times \{ \frac{{10/\ln 10}}{{\sqrt {2\pi {\sigma ^2}} \Omega }}\exp [ - \frac{{(10{{\log }_{10}}\Omega  - \mu )}}{{2{\sigma ^2}}}]\} d\Omega ,\;\;\gamma  \ge 0,\\
\end{array}
\end{equation}
where $\Omega$ is the average power which is treated as a random variable, and $m$ is the parameter of Nakagami-m fading.
$\sigma(dB)$  and $\mu(dB)$ are the mean and standard deviation of $10{\log _{10}}\Omega $, respectively. Then the outage probability is obtained as
\begin{equation}\label{eq-17}
\begin{array}{l}
{P_{out}} = \int_0^{{\gamma _{th}}} {\int_0^\infty  {\frac{{{m^m}{\gamma ^{m - 1}}}}{{{\Omega ^m}\Gamma (m)}}} \exp ( - \frac{{m\gamma }}{\Omega })} \\
\;\;\;\; \times \{ \frac{{10/\ln 10}}{{\sqrt {2\pi {\sigma ^2}} \Omega }}\exp [ - \frac{{(10{{\log }_{10}}\Omega  - \mu )}}{{2{\sigma ^2}}}]\} d\Omega d\gamma.
\end{array}
\end{equation}
In formula (\ref{eq-15}), the instantaneous SNR $\gamma$ and the threshold of SNR can be presented as
\begin{equation}\label{eq-18}
\begin{array}{l}
\gamma  = \frac{{{\alpha ^2}{E_s}}}{{{N_0}}} = \frac{{{\alpha ^2}{P_{R}^b}{T_s}}}{{N/B}} = \frac{{{\alpha ^2}{P_{R}^b}}}{N}\;\;\;\;\;\;\;\;\;(a),\\
{\gamma _{th}} = \frac{{{\alpha ^2}{P_{th}}}}{N} \;\;\;\;\;\;\;\;\;\;\;\;\;\;\;\;\;\;\;\;\;\;\;\;\;\;\;\;\;\;\;\;\;\;\;(b),
\end{array}
\end{equation}
where $\alpha ^2$ is the fading power, $E_s$ indicates the energy per symbol and $N_0(W/Hz)$ is the one-sided power spectral density at the receiver. $P_{rb}$ is the received signal power and $N$ is the noise power at the receiver. $T_s$ is the period of the signal and $B_s$ is the bandwidth of the noise, and it satisfies $B_s = \frac{1}{{{T_s}}}$, so we can get the final expression in formula (\ref{eq-18}-a).

Then by substituting formula (\ref{eq-18}-a) into (\ref{eq-17}), we can get the outage probability varying with the received signal power $P_{rb}$ which is the signal power before experiencing fading. Thus, the threshold of this power can be obtained by analyzing the outage probability shown in formula (\ref{eq-17}), that is, the threshold of $P_{th}$ in formula (\ref{eq-14}) is obtained. Finally, to calculate the threshold of the transmission power, we can analyze the formula (\ref{eq-14}) since the value range of the received power $P_{th}$ is known now.

\section{Numerical Simulations }
In this section, simulations are performed for both optimal timeslot allocation schemes proposed in Section II. For comparison, the relevant simulation parameters are set as the same, $T = 10s$, the consumption energy of the sensor is ${P_{D}} =  - 30dBm$, and the circuit power of backscatter module is  ${P_C} = -40dBm$. The consumption energy of each sample is $e_s=-33dBm$. The path-loss exponent for both communication links are $\varsigma=2$. The sampling rate of each signal is $f_s=1000Hz$, and the noise power at the backscatter receiver is given as $N=-40dBm$. And then simulations are performed to find the detection threshold of spectrum sensing for backscatter communication.

\subsection{Optimal Scheme with spectrum sensing}
\begin{figure}[t]
\begin{center}
\includegraphics[width=3.3in]{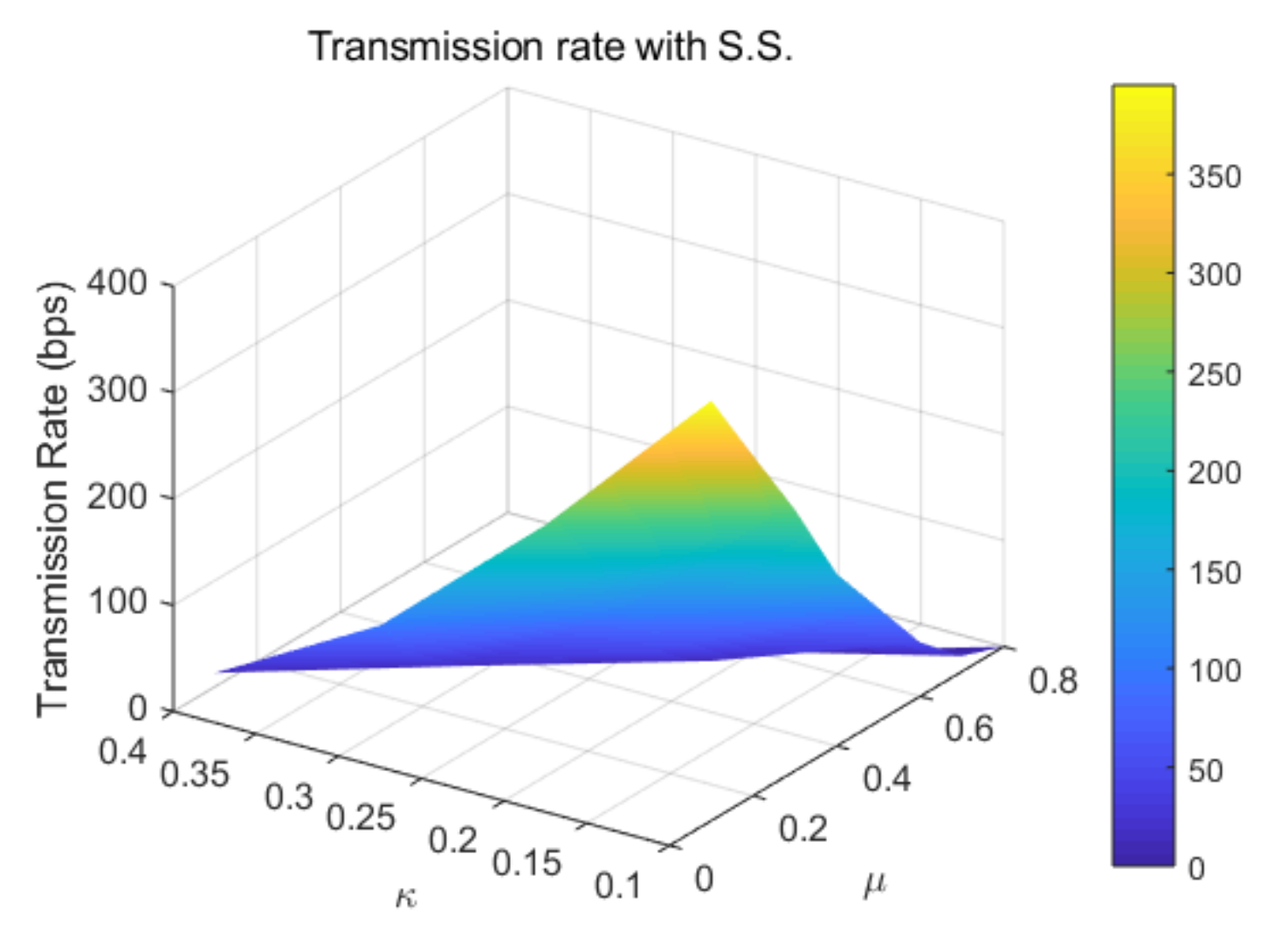}    
\caption{The transmission rate with spectrum sensing }  
\label{fig5}                                 
\end{center}                                 
\end{figure}

\begin{figure}[t]
\begin{center}
\includegraphics[width=3.3in]{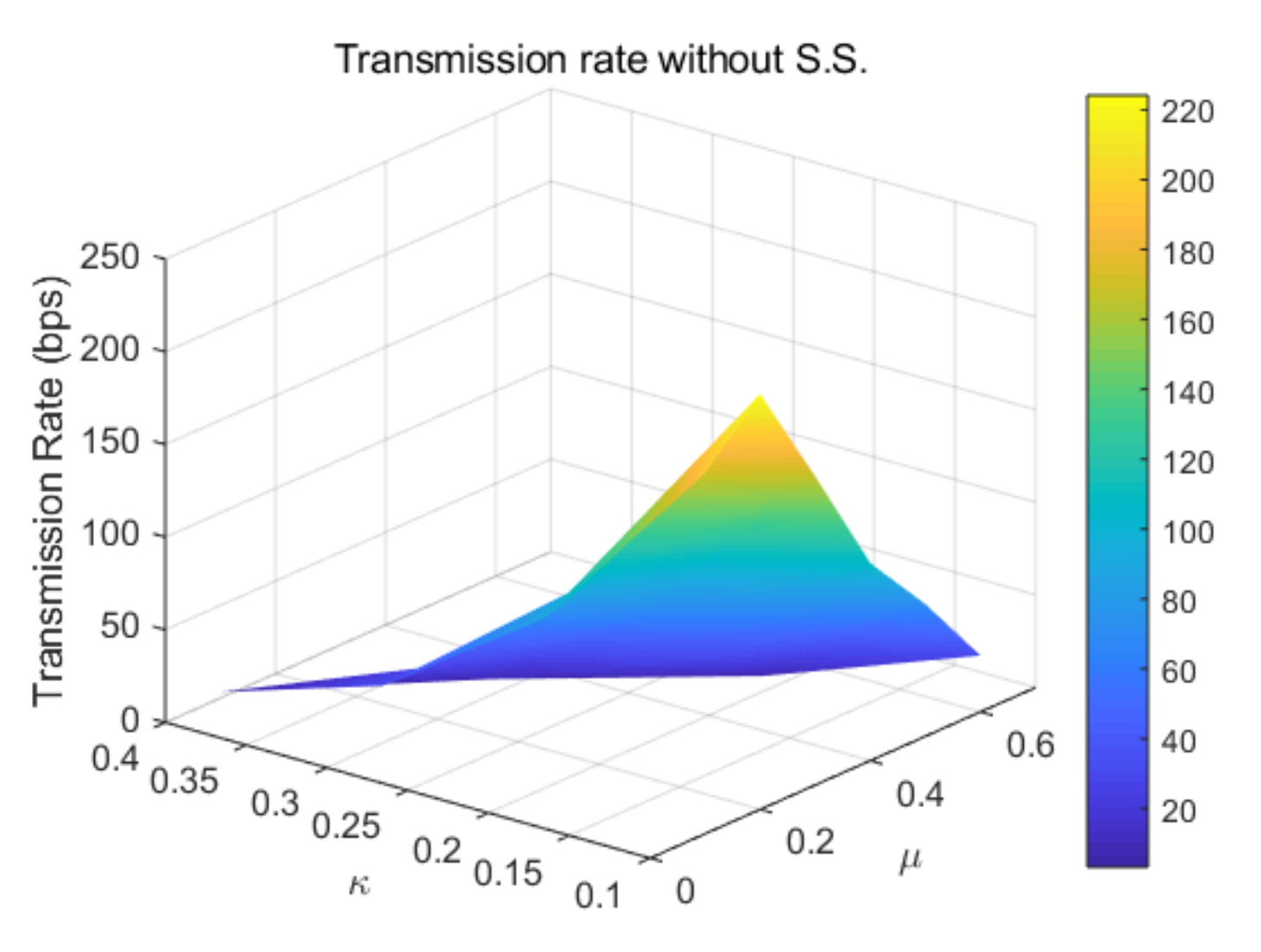}    
\caption{The transmission rate without spectrum sensing}  
\label{fig6}                                 
\end{center}                                 
\end{figure}

In this section, we detect the signals from a single frequency which is set as 562MHz. By solving the optimization problem OP1, the optimal transmission rate of backscatter communication is obtained with optimal timeslot allocation parameters, which is shown in Fig.~\ref{fig5}. In contrast, Fig.~\ref{fig6} plots the transmission rate of the backscatter communication without adopting the energy detection technique. In this case, it means that the energy harvesting efficiency $\eta$ and the reflection coefficient $\beta$ of backscatter communication are smaller than 1, which we set as 0.5. This is because the received signals that don't satisfy the energy harvesting or backscatter communication condition will be eliminated. And the optimal values of transmission rates and the corresponding optimal timeslot allocation parameters are given in Table 1(a).

Comparing Fig.~\ref{fig5} to Fig.~\ref{fig6}, we can see that the transmission rate is improved. This is because we use a pre-defined threshold to filter the received signals for energy harvesting and backscatter communication. So in this case, $\eta$ and $\beta$ are equal to 1, which means all the filtered signals arrived in the energy harvesting module and the backscatter module can be harvested and modulated successfully. We also illustrate the optimal value ranges of timeslot allocation parameters $\kappa$ and $\mu$ in these two scenarios.

\begin{table}
\caption{Optimal Values of both Schemes}
\centering
\subtable[ Optimal Values with spectrum sensing ]{
       \begin{tabular}{|c|c|c|c|c|}
        \hline
       & ${R_n}(bps)$& $\kappa$ & $\mu$ &$\tau$\\
        \hline
        s.s. &395 & 0.11  & 0.11& 0.78\\
        \hline
        no s.s. & 224 & 0.11 & 0.21& 0.68\\
        \hline

       \end{tabular}
       \label{tab:firsttable}
}
\qquad
\subtable[Optimal Values with compressive sensing]{
       \begin{tabular}{|p{0.9cm}<{\centering}|p{0.9cm}<{\centering}|p{0.9cm}<{\centering}|p{0.9cm}<{\centering}|}
       \hline
     & ${R_w}$& $\alpha$ & $\gamma$\\
        \hline
        c.s. & 3864  & 0.11  & 0.11\\
        \hline
         no c.s. & 2694  & 0.21  & 0.11\\
         \hline
       \end{tabular}
       \label{tab:secondtable}
}
\end{table}
\subsection{Optimal Scheme with compressive sensing}
To improve the spectrum sensing efficiency, compressive sensing is used to detect the wideband TV signals.
There are 40 channels ranging from 470 to 790 MHz, so the number of detected wideband signals is 40. And the sparsity of the signal for compressive sensing is set as $K= 75\%$.

\begin{figure}[t]
\begin{center}
\includegraphics[width=3.3in]{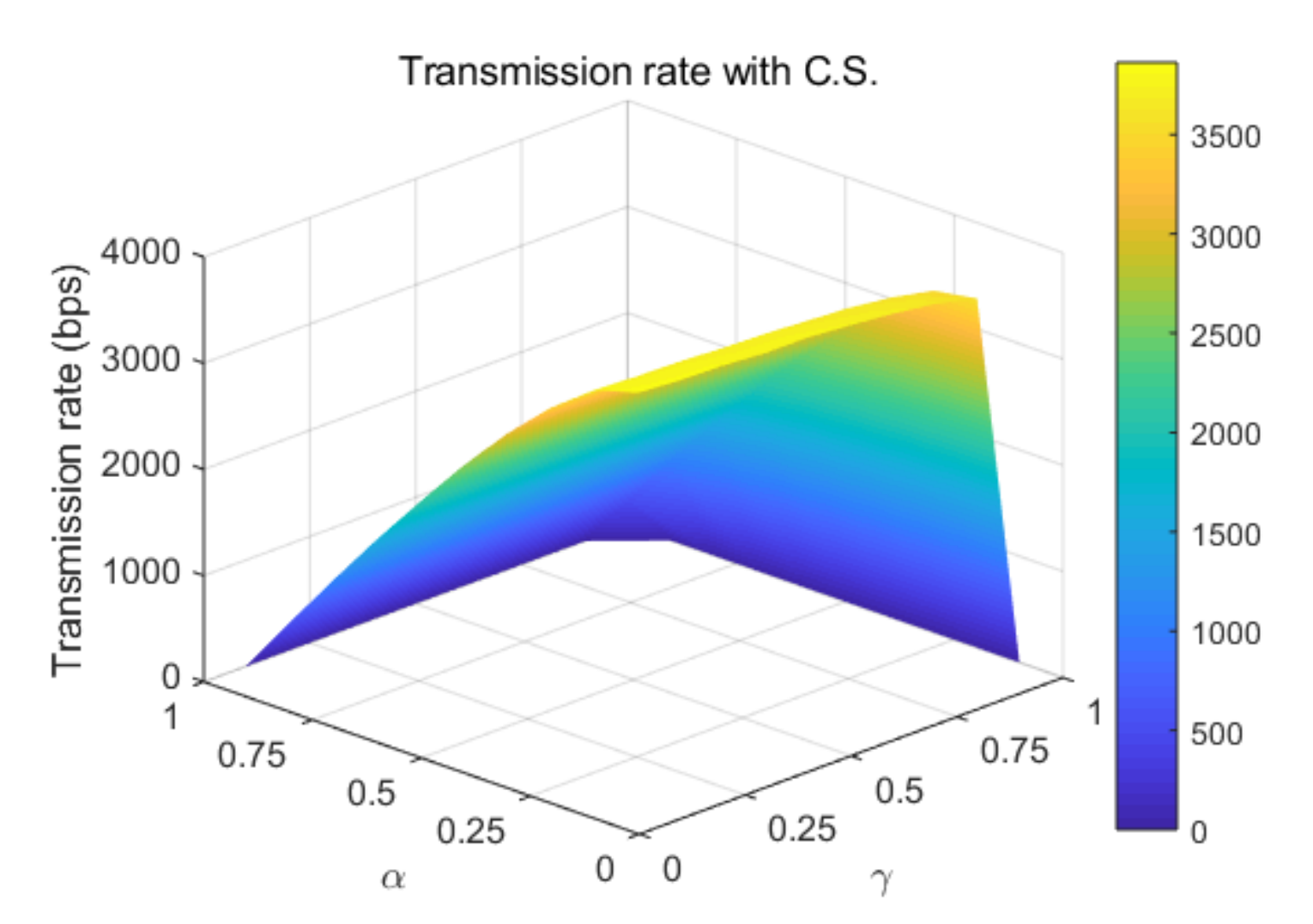}    
\caption{The transmission rate with compressive sensing}  
\label{fig7}                                 
\end{center}                                 
\end{figure}

\begin{figure}[t]
\begin{center}
\includegraphics[width=3.3in]{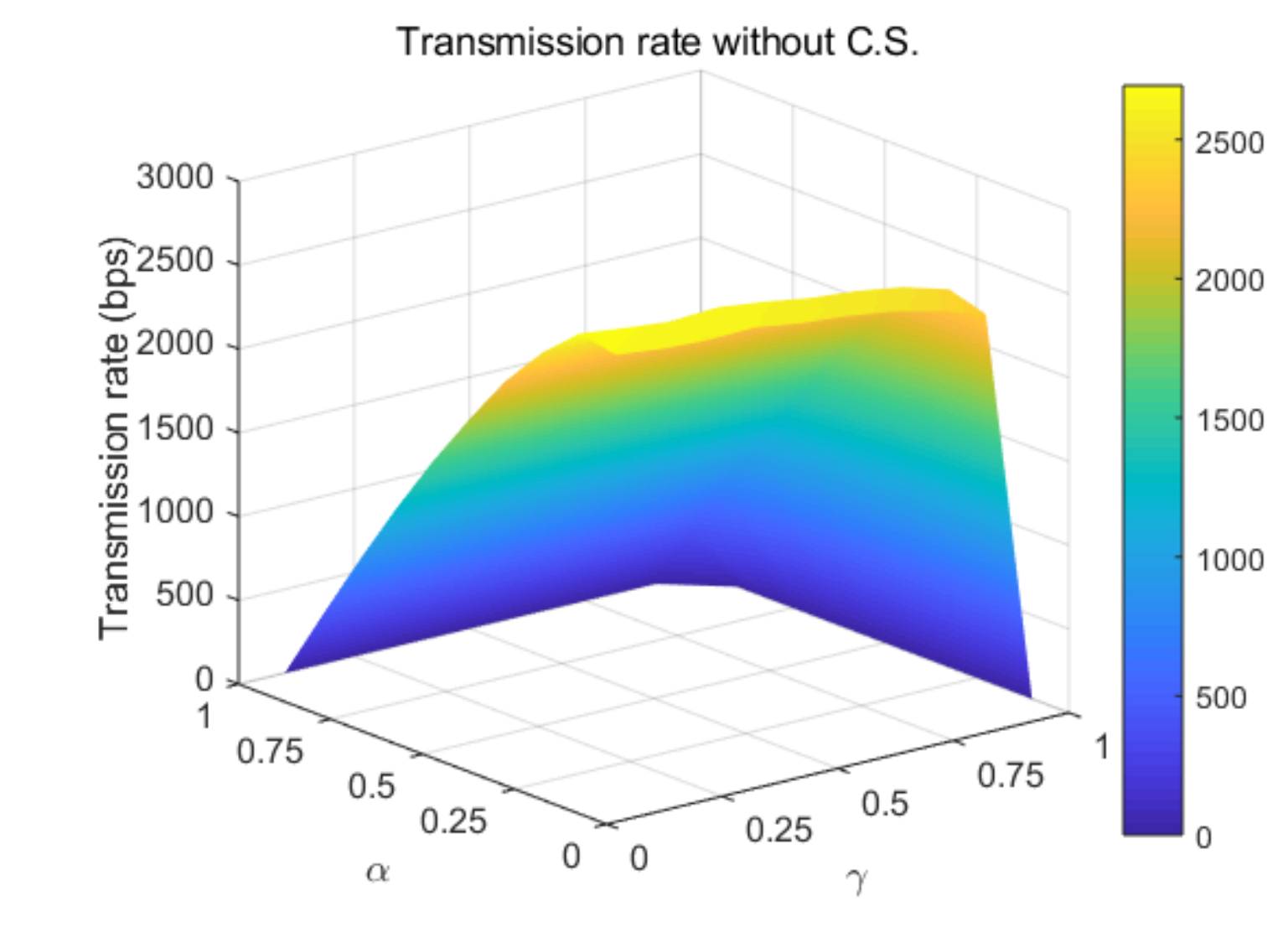}    
\caption{The transmission rate without compressive sensing}  
\label{fig8}                                 
\end{center}                                 
\end{figure}
For comparison, leveraging this wideband signal without compressive sensing is also performed. Similarly, the energy harvesting efficiency $\eta$ and the reflection coefficient $\beta$ are set as 0.5 in this scenario. In this scheme, both the time scheduling parameter and the power allocation ratio have effects on the transmission rate. The obtained optimal values of them and the corresponding transmission rates in both scenarios are given in Table 1(b).

Comparing Fig.~\ref{fig7} to Fig.~\ref{fig8}, we can see that the transmission rate with compressive sensing is improved. Likewise, this is because all the RF signals arrived in the energy harvesting module and the backscatter module are filtered by using the energy detection technique. The performance of the transmission rate is obviously improved with compressive sensing, which is shown in Fig.~\ref{fig5} and Fig.~\ref{fig7}. The reason is that the wideband multi-frequency signal can be detected at the same time with compressive sensing, which increases the incident power of energy harvesting and backscatter communication.

\begin{figure}[t]
\begin{center}
\includegraphics[width=3in]{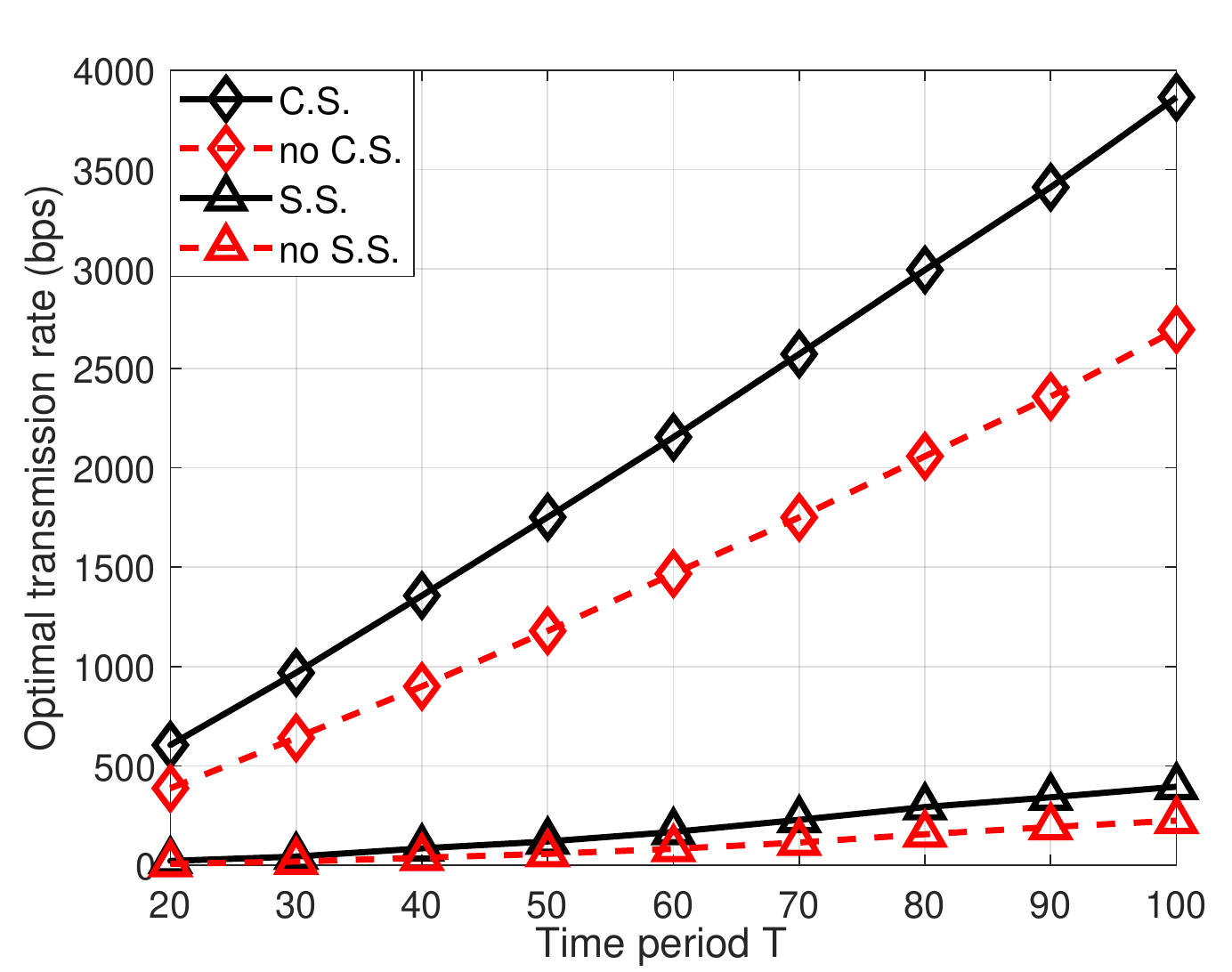}    
\caption{The transmission rate with increasing network operation time }  
\label{increasing_rate}                                 
\end{center}                                 
\end{figure}

In addition, as shown in Fig.~\ref{increasing_rate}, we can easily see that the optimal scheme with spectrum sensing technique has higher transmission rate, and by using compressive sensing, the performance is improved obviously. Fig.~\ref{increasing_rate} also illustrates that the superiority of the scheme using spectrum sensing becomes more obvious with the increasing of the operating time of the IoT network, and it is the same with the superiority of the time scheduling scheme with compressive sensing.

\subsection{Detection Threshold for Backscatter Communication}
In this section, simulations are performed to obtain the detection power threshold of spectrum sensing module for backscatter communication. It is calculated by the reverse derive method, firstly, we analyze the outage probability of backscatter communication suffering composite log-normal shadowing and Nakagami-m fading through formula (\ref{eq-17}). Then by analyzing the outage probability of backscatter communication using path loss model through formula (\ref{eq-14}), we can obtain the detection threshold of spectrum sensing for backscatter communication. The simulation parameters are set as: the shadowing scenario is set as average shadowing, where $\mu=-0.115$ and $\sigma=0.161$, and the channel fading amplitude $\alpha=0.7$\cite{add_3coulson1998statistical}.

\begin{figure}[t]
\begin{center}
\includegraphics[width=3in]{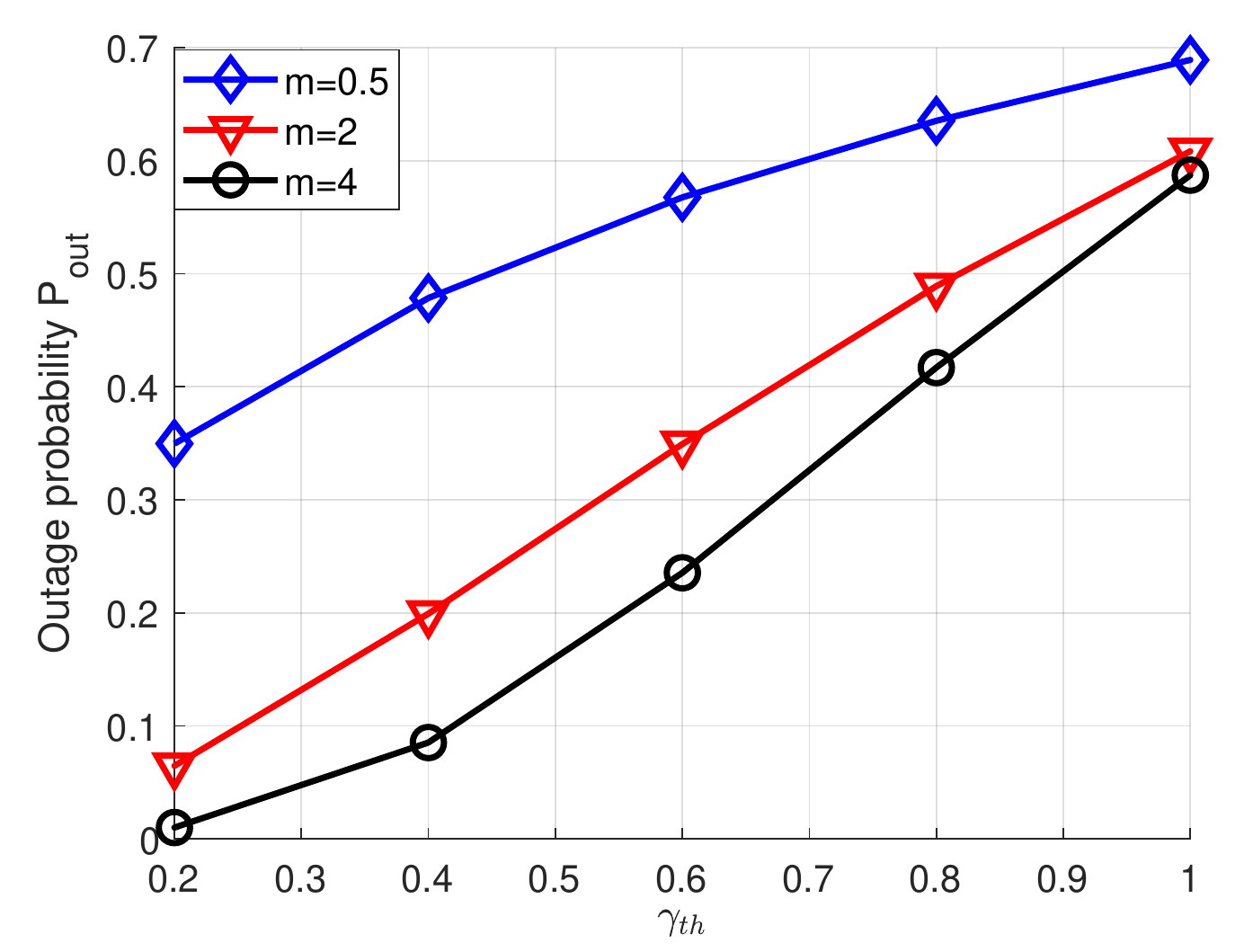}    
\caption{outage probability $P_{out}$ vs SNR threshold $\gamma_{th}$}  
\label{pout_gamma}                                 
\end{center}                                 
\end{figure}

\begin{figure}[t]
\begin{center}
\includegraphics[width=3in]{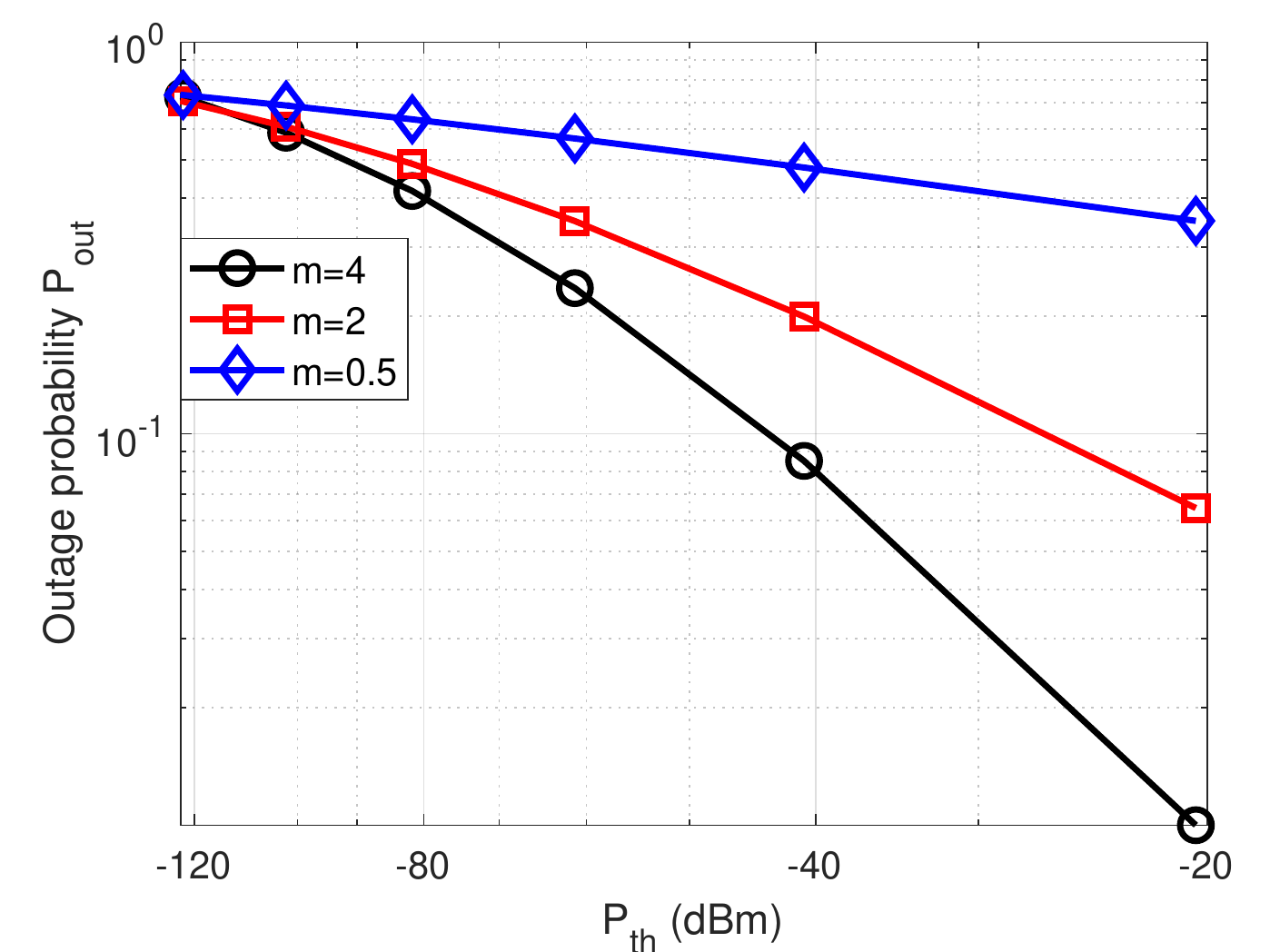}    
\caption{outage probability $P_{out}$ vs received power $P_{th}$}  
\label{pout_pth}                                 
\end{center}                                 
\end{figure}

As shown in Fig.~\ref{pout_gamma}, the outage probability is increasing with the increasing of the SNR threshold $\gamma_{th}$ at the receiver. According to the definition of the outage probability, it has higher possibility to suffer outage with higher SNR requirements. And in formula (\ref{eq-18}), the SNR threshold $\gamma_{th}$ and the received power $P_{th}$ are in linear relation, so the outage probability varying with $P_{th}$ is easily plotted in Fig.\ref{pout_pth}. It shows that the $P_{out}$ decreases with the received power $P_{th}$ increasing under different Nakagami-m fading parameters. And with the increasing of the fading parameter $m$, the $P_{out}$ is decreased. This is because the Nakagami-m fading converges to a non-fading AWGN channel with $m$ increasing to $ + \infty $.


\begin{figure}[t]
\begin{center}
\includegraphics[width=3in]{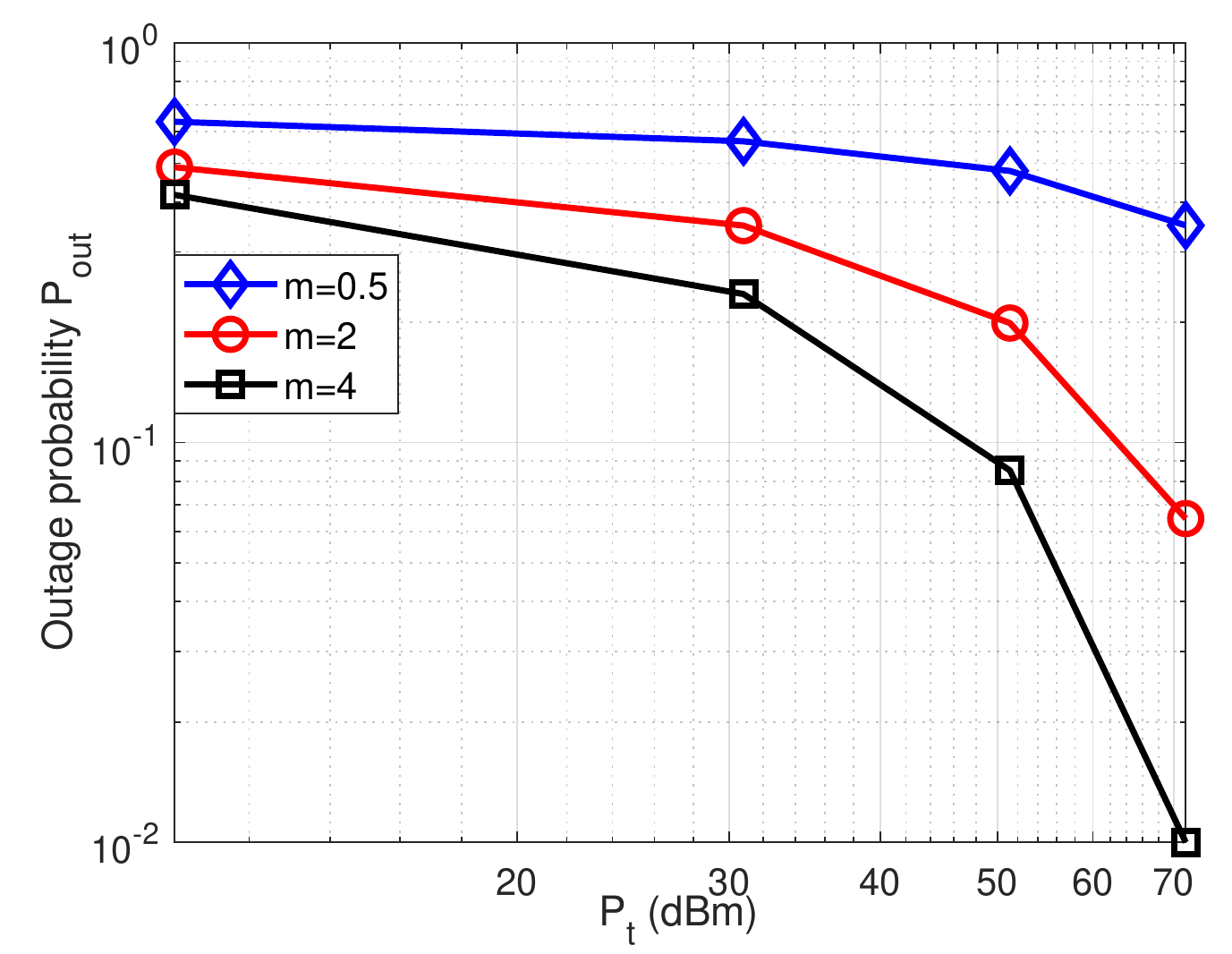}    
\caption{outage probability $P_{out}$ vs transmission power $P_{t}$}  
\label{pout_pt}                                 
\end{center}                                 
\end{figure}

\begin{figure}[t]
\begin{center}
\includegraphics[width=3in]{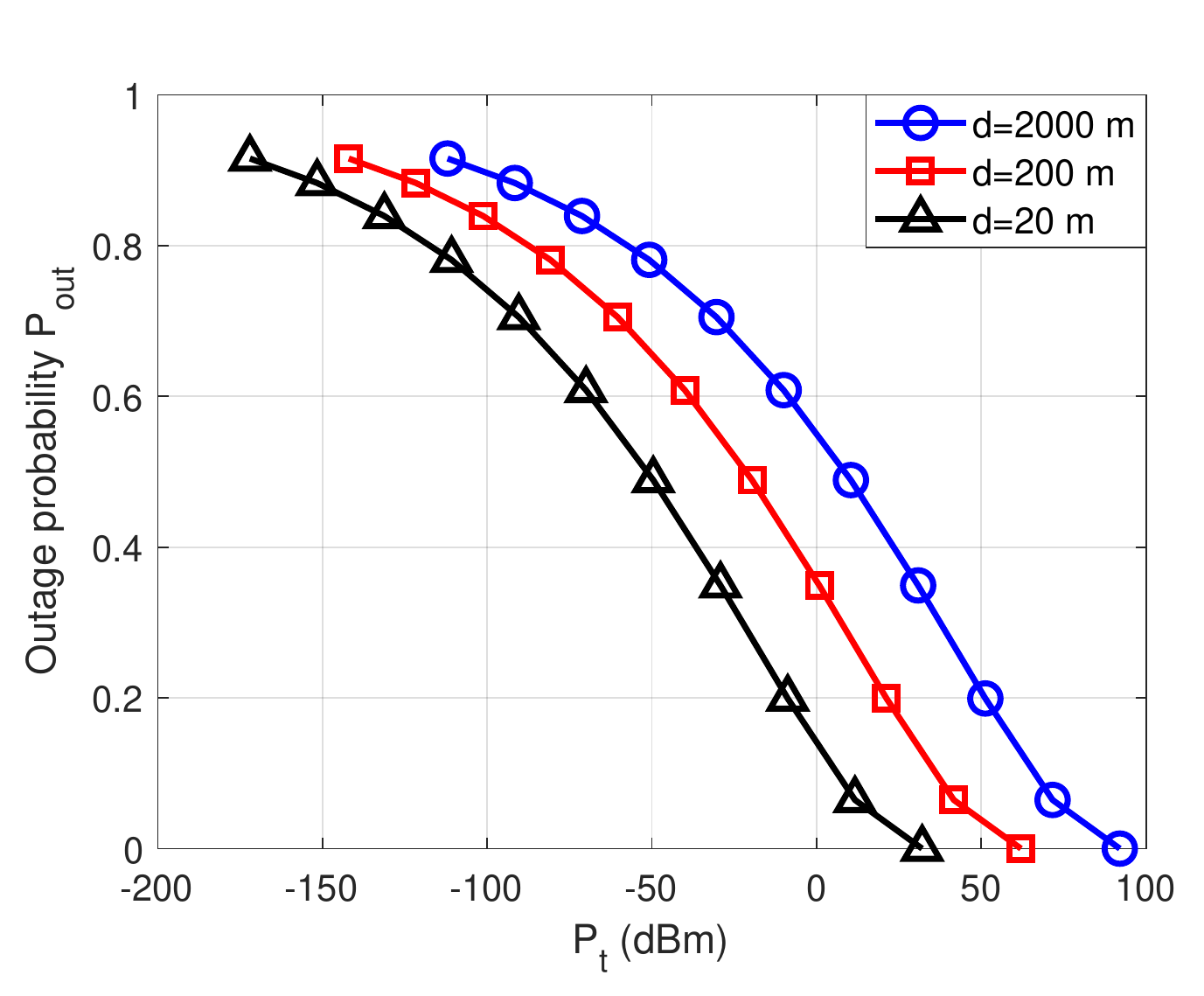}    
\caption{outage probability $P_{out}$ with different distances}  
\label{pout_d}                                 
\end{center}                                 
\end{figure}

Afterwards, by analyzing the path loss model between the transmission power and the received power shown in formula (\ref{eq-14}), we obtain the outage probability varying with the transmission power as shown in Fig.~\ref{pout_pt}, that is, the threshold of spectrum sensing for backscatter communication can be determined by controlling the outage probability. And Fig.~\ref{pout_d} illustrates the analogous performance of $P_{out}$ with $P_{th}$ over different distances between the backscatter device and the receiver. It is obvious that the communication link tends to be interrupted with the increasing of the communication distance.

\section{Conclusion}
In this paper, the optimal time scheduling scheme based on spectrum sensing technique has been proposed to optimize the time scheduling of spectrum sensing module, the energy harvesting module and the ABCom module at the IoT node. By maximizing the transmission rate of backscatter communication, optimal time scheduling parameters and the optimal power allocation ratio have been obtained. Simulations demonstrate that larger transmission rates have been achieved when using spectrum sensing techniques, and compressive sensing has better performance. The superiorities become more obvious with the increasing of the network operation time.
Additionally, we obtained the detection threshold of spectrum sensing for enabling backscatter communication by analyzing the outage probability of ambient backscatter communication.


\bibliographystyle{IEEEtran}
\bibliography{IEEEabrv,liuxiaolanreference}

\end{document}